# Quantum Transitions in Terms of non-Adiabatic Dressed States

**I. G. Koprinkov**
*Department of Applied Physics, Technical University of Sofia, 8 Kl. Ochridski Blvd., 1756 Sofia, Bulgaria*
*e-mail: igk@tu-sofia.bg*

Nonadiabatic dressed states of a quantum system interacting with an external electromagnetic field and the environment are presented. The relevant matrix elements within the specified states are found. A closed form expression of the quantum transitions based on the nonadiabatic dressed states is proposed for the first time.

PACS numbers: 03.65.-w , 42.50.Ct

**1. Introduction**

Quantum transitions are among the most fundamental physical phenomena, leading to non-trivial changes in the state of the quantum systems. The problem of quantum transitions becomes a crossing point for a number of physical concepts and approaches. One of the basic reference points to the quantum transitions is *the adiabatic theorem of quantum mechanics* [1]. According to it, if the Hamiltonian changes infinitely slowly, *i.e.*, adiabatically, the quantum system gradually rearranges its configuration thus remaining in the same eigenstate of the instantaneous Hamiltonian, and no transitions to other states will occur. If the Hamiltonian changes rapidly, the quantum system does not manage to rearrange its configuration and the final state may differ from the initial state once the perturbation is removed, *i.e.*, transitions to other states may occur. Such transitions are called nonadiabatic transitions. Another basic concept involved in quantum transitions is *the Born-Oppenheimer adiabatic approximation* [2]. It takes place for quantum systems with internal degrees of freedom (molecules, condensed phase, etc.) and rests on the large difference between the electron and the nucleus masses. In the lowest order of approximation, the total wave function can be factorized into (adiabatic) electronic and nuclear states. The higher orders of approximation introduce nonadiabatic factors that couple the adiabatic states and lead to nonadiabatic transitions. The nonadiabatic transitions in this case arise from a breakdown of the Born-Oppenheimer approximation that allows to gain energy from the nuclear motion. *The formal theory of quantum transitions* is based on the concept of perturbation as a reason for quantum transitions [3-5]. The adiabatic theorem specifies that not all but the nonadiabatic perturbations lead in fact to quantum transitions. The above concepts outline the physical factors of the quantum transitions: *(i) external nonadiabatic factors* from "regular" electromagnetic fields and relaxation processes (damping) due to stochastic fields from the environment (zero-point vacuum fluctuations, collisions, etc.) [6-8], and *(ii) internal nonadiabatic factors* leading to violation of the Born-Oppenheimer approximation [9]. The problem of quantum transitions, dating back to the early days of quantum mechanics, was solved first for the one-dimensional two-level case of diabatic (crossing) [10, 11] and adiabatic (noncrossing) states [12]. Later on, it was extended to multidimensional and multilevel systems [13, 14]. The problem is usually treated within the bare states (BS), *i.e.*, the eigenstates of the non-perturbed Hamiltonian $\hat{H}_0$ of the quantum system. The states of the "atom"-field system, known as dressed states, are also used [15, 6-8]. Both kinds of states do not involve nonadiabatic factors and, from such a point of view, can be considered as adiabatic states. Therefore, the dressed states will be named adiabatic dressed states (ADS). They are eigenstates of the adiabatic Hamiltonian $\hat{H}_A = \hat{H}_0 + \hat{H}_A'$, where $\hat{H}_A'$ is adiabatic perturbation. Here, we consider for the first time [16] quantum transitions based on nonadiabatic dressed states (NADS) [17] that include explicitly nonadiabatic factors from the field and the damping. The relevant matrix elements of the NADS are found. A closed form expression of the quantum transitions within the NADS is obtained.

**2. Non-adiabatic Dressed States**

The NADS are derived from an analytic solution of the Schrödinger equation $\hat{H}|\Psi(\vec{r},t)\rangle = i\hbar\partial_t|\Psi(\vec{r},t)\rangle$ for an open quantum system of Hamiltonian $\hat{H}$, Eq.(1) (in standard notations), subject to a "regular" interaction $\hat{H}'$ with an external electromagnetic field $E(t)$ (considered, in general, as nonadiabatic) and to a stochastic type of interaction $\hat{H}_D$ with the environment of damping rates $\gamma_j$ (that makes the Hamiltonian non-Hermitian), Fig.1:

$$\hat{H} = \hat{H}_0 + \hat{H}' + \hat{H}_D = \sum_{j=1}^{2}\hbar\omega_j|j\rangle\langle j| - \mu E(t)(|1\rangle\langle 2| + h.c.) - i\hbar\sum_{j=1}^{2}(\gamma_j/2)|j\rangle\langle j| \quad . \quad (1)$$

The states $|j\rangle$, ground $|1\rangle \equiv |g\rangle$ and excited $|2\rangle \equiv |e\rangle$, are the original BS of the quantum system. The electromagnetic field is presented in terms of the carrier-envelop concept $E(t) = (1/2)E_o(t)[\exp\{i(\omega t + \varphi(t))\} + c.c.]$, where $E_o(t)$ and $\varphi(t)$ are the envelope and the phase of the field, respectively, considered as arbitrary functions of time, subject to a generalized adiabatic condition only [17], $\omega$ is the carrier frequency, and *c.c.* means complex conjugate. It holds for a broad range of fields down to single-cycle ultrashort pulses [18], having few femtoseconds pulse duration in the optical range. It allows to involve the ultrafast phenomena that are strongly influenced by the nonadiabatic effects.

Following the procedure described in [17], the following solution of the Schrödinger equation for the specified problem at ground, Eq. (2), and excited, Eq. (3), state initial conditions is obtained

$$|\Psi(\vec{r},t)\rangle = (\tilde{\Omega}')^{-1/2} \left\{ (\tilde{\Omega}'_0)^{-1/2} \tilde{\Lambda}'_{10} COS^{-1}(\theta/2)|\tilde{G}'\rangle + (\tilde{\Omega}')^{-1/2} \tilde{\Lambda}'_{20} SIN^{-1}(\theta/2)|\tilde{E}'\rangle \right\} \qquad (2)$$

$$|\Psi(\vec{r},t)\rangle = (\tilde{\Omega}')^{-1/2} \left\{ 2^{-1}(\tilde{\Omega}'_0)^{-1/2} \Omega_0 COS^{-1}(\theta/2)|\tilde{G}'\rangle + 2^{-1}(\tilde{\Omega}'_0)^{-1/2} \Omega_0 SIN^{-1}(\theta/2)|\tilde{E}'\rangle \right\} \qquad (3)$$

$|\tilde{G}'\rangle$ and $|\tilde{E}'\rangle$ are ground and excited NADS, respectively, that consist of real ("r") and virtual ("v") components

$$|\tilde{G}'\rangle = COS(\theta/2)|\tilde{G}'\rangle_r + SIN(\theta/2)|\tilde{G}'\rangle_v \quad , \quad |\tilde{E}'\rangle = COS(\theta/2)|\tilde{E}'\rangle_r - SIN(\theta/2)|\tilde{E}'\rangle_v \qquad (4)$$

The real and the virtual components of the NADS (at ground state initial conditions) can be expressed as

$$|\tilde{G}'\rangle_r = |g\rangle \exp(-i\int_0^t \tilde{\omega}'_G dt') \qquad |\tilde{E}'\rangle_r = |e\rangle \exp[-i\int_0^t \tilde{\omega}'_E dt' - i\varphi(t)]$$
$$|\tilde{G}'\rangle_v = |e\rangle \exp[-i\int_0^t (\tilde{\omega}'_G + \omega)dt' - i\varphi(t)] \quad , \quad |\tilde{E}'\rangle_v = |g\rangle \exp(-i\int_0^t (\tilde{\omega}'_E - \omega)dt') \qquad (5)$$

Quantities $COS(\theta/2) = \sqrt{\tilde{\Lambda}'_1/\tilde{\Omega}'}$ and $SIN(\theta/2) = sgn(\Delta\omega)\sqrt{-\tilde{\Lambda}'_2/\tilde{\Omega}'}$ are complex functions that determine the partial contribution of the real and virtual components of the NADS, $\tilde{\Omega}' = sgn(\Delta\omega)[\Omega^2 + \Delta\tilde{\omega}'^2 - i2\partial_t \Delta\tilde{\omega}']^{1/2}$ is nonadiabatic Rabi frequency, $\Delta\tilde{\omega}' = \Delta\omega - i(\gamma_g + \gamma_e)/2 - (\partial_t\varphi - i\Omega^{-1}\partial_t\Omega)$ is nonadiabatic frequency detuning, $\Lambda_1 = (\Delta\tilde{\omega}' + \tilde{\Omega}')/2$, $\Lambda_2 = (\Delta\tilde{\omega}' - \tilde{\Omega}')/2$ and $\tilde{\Lambda}'_j = \Lambda_j - i(2\tilde{\Omega}')^{-1}\partial_t\tilde{\Omega}'$ are generalizations of the respective quantities from the adiabatic case [17], $\tilde{\omega}'_G = \omega_g + \Lambda_2$ and $\tilde{\omega}'_E = \omega_e - \Lambda_2 - i(\gamma_g + \gamma_e)/2 - (\partial_t\varphi - i\Omega^{-1}\partial_t\Omega)$ are nonadiabatic frequencies (energies) of the ground and the excited NADS, respectively, and $\Omega(t) = \mu E_o(t)/\hbar$ and $\Delta\omega = \omega_e - \omega_g - \omega$ are the usual Rabi frequency and frequency detuning. The subscript "0" means the initial value of the quantity. The damping ($\gamma_g$, $\gamma_e$) and the time derivatives of the field (amplitude/Rabi frequency $\Omega^{-1}\partial_t\Omega$ and phase $\partial_t\varphi$) are *nonadiabatic factors*. The above solution is *non-perturbative* and *non-adiabatic* simultaneously. The NADS ($|\tilde{G}'\rangle$, $|\tilde{E}'\rangle$) represent a generalization of the BS ($|g\rangle$, $|e\rangle$) and the ADS ($|G\rangle$, $|E\rangle$): BS + (*adiabatic*) *field* → ADS ; ADS + *nonadiabatic factors* → NADS .

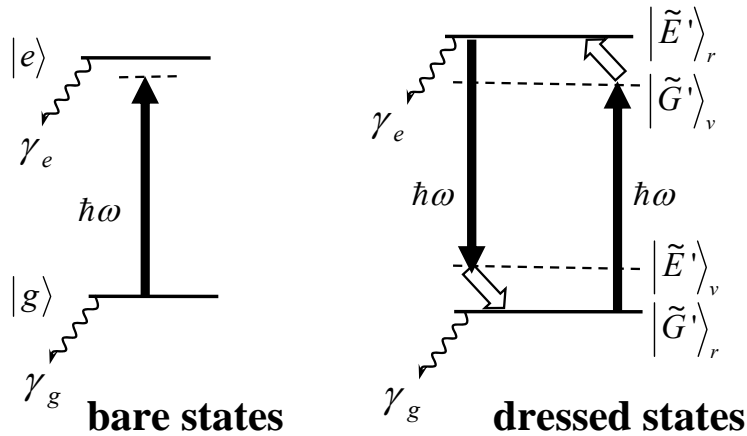

**Fig.1:** Bare and dressed states of a quantum system. The full lines represent radiative transitions, the hollow lines represent nonadiabatic transitions, and the wavy lines represent relaxation processes.

**3. Quantum Transitions in Terms of the Non-Adiabatic Dressed States**

The quantum mechanical problems that are usually performed within the BS now can be reconsidered within the NADS. The NADS and their parameters include explicitly the nonadiabatic factors from the field and the damping. This will automatically incorporate the nonadiabatic factors in the phenomena under investigation.

The BS, the ADS, and the NADS derive continuously one from the other in this or in a reverse order switching on/off the field and the nonadiabatic factors, respectively. The nonadiabatic factors can be changed continuously by varying the parameters of the field and/or the damping rates, assuming the latter can be done continuously from zero (a closed quantum system) to a given non-zero value (an open quantum system). In view of that, it is reasonable to use the same, Hermitian, inner product for the NADS as it is in the case of the BS and the ADS. Some of the main matrix elements between the NADS, relevant to the problem under investigation, can thus be obtained:

$$\langle '\tilde{G}|\tilde{G}'\rangle = \left[|SIN(\theta/2)|^2 + |COS(\theta/2)|^2\right]\exp\left(2\int_0^t \mathrm{Im}\,\tilde{\omega}'_G\, dt'\right) \qquad (6)$$
$$= \left[|SIN(\theta/2)|^2 + |COS(\theta/2)|^2\right]\exp\left\{-(\gamma_1/2+\gamma_2/2)t + \int_0^t (\Omega^{-1}\partial_t\Omega - \mathrm{Im}\,\tilde{\Omega}')dt'\right\}$$

$$\langle '\tilde{E}|\tilde{E}'\rangle = \left[|SIN(\theta/2)|^2 + |COS(\theta/2)|^2\right]\exp\left(2\int_0^t \mathrm{Im}\,\tilde{\omega}'_E\, dt'\right) \qquad (7)$$
$$= \left[|SIN(\theta/2)|^2 + |COS(\theta/2)|^2\right]\exp\left\{-(\gamma_1/2+\gamma_2/2)t + \int_0^t (\Omega^{-1}\partial_t\Omega + \mathrm{Im}\,\tilde{\Omega}')dt'\right\}$$

$$\langle '\tilde{E}|\tilde{G}'\rangle = \left[SIN(\theta/2)(COS(\theta/2))^* - (SIN(\theta/2))^* COS(\theta/2)\right]\exp\left\{i\int_0^t \left[\tilde{\omega}'^*_E - \tilde{\omega}'_G - \omega\right] dt'\right\} \qquad (8)$$
$$= \left[SIN(\theta/2)(COS(\theta/2))^* - (SIN(\theta/2))^* COS(\theta/2)\right]\exp\left\{-(\gamma_1/2+\gamma_2/2)t + \int_0^t (\Omega^{-1}\partial_t\Omega + i\,\mathrm{Re}\,\tilde{\Omega}')dt'\right\}$$

The first line in Eqs.(6)-(8) is a concise form of the matrix element while the second one gives a more expanded form. The following features take place for the matrix elements: $\langle '\tilde{G}|\tilde{G}'\rangle^* = \langle '\tilde{G}|\tilde{G}'\rangle$, $\langle '\tilde{E}|\tilde{E}'\rangle^* = \langle '\tilde{E}|\tilde{E}'\rangle$, $\langle '\tilde{G}|\tilde{E}'\rangle^* = \langle '\tilde{E}|\tilde{G}'\rangle$, where " * " means complex conjugate. The matrix elements $\langle '\tilde{G}|\tilde{G}'\rangle$ and $\langle '\tilde{E}|\tilde{E}'\rangle$ are real and positive definite, which makes the norm of the NADS, $\||\tilde{G}'\rangle\| \equiv (\langle '\tilde{G}|\tilde{G}'\rangle)^{1/2}$, $\||\tilde{E}'\rangle\| \equiv (\langle '\tilde{E}|\tilde{E}'\rangle)^{1/2}$, meaningful.

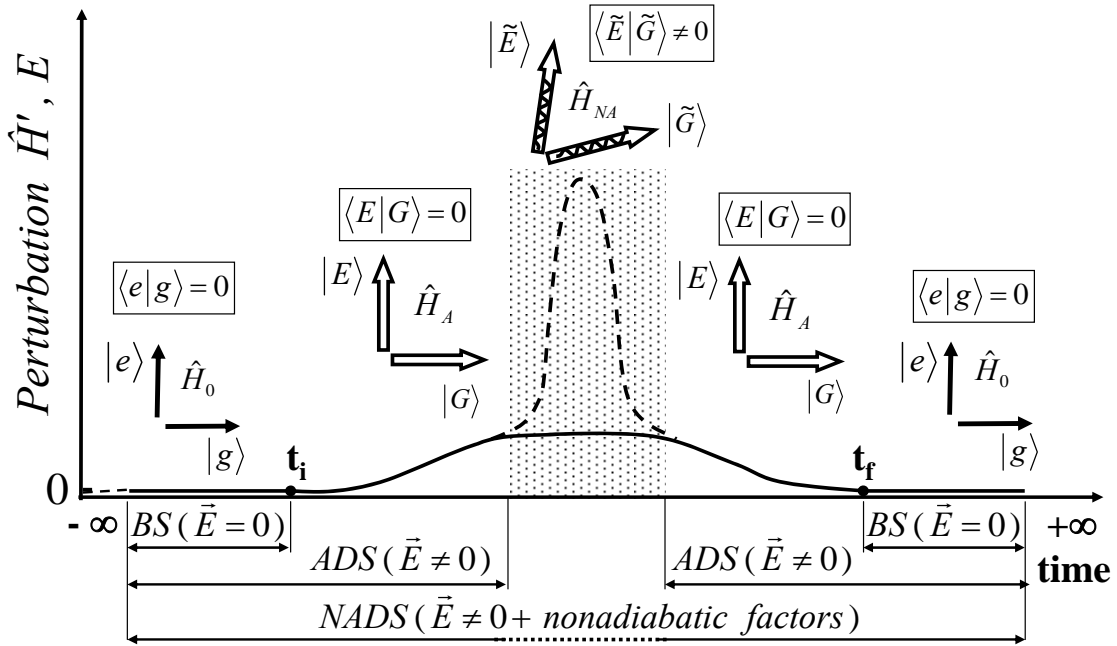

**Fig.2:** BS, ADS, NADS, switch-on/off $t_i/t_f$ time of the perturbation, and orthogonality of the states schematic view.

The BS, the ADS, the NADS, the switch on/off time $t_i/t_f$ of the perturbation, and the properties of orthogonality of the states are shown schematically in Fig.2. The adiabatic perturbation/field $\hat{H}'/\vec{E}$ is shown with a full line while the non-adiabatic perturbation is shown with a broken line. Each BS gives rise to corresponding ADS or NADS depending on the type of the perturbation $\hat{H}'$, adiabatic or nonadiabatic, respectively. Thus, the BS, the ADS, and the NADS form a complete set of linearly independent state vectors. The BS and the ADS are orthogonal, $\langle e|g\rangle = 0$, $\langle E|G\rangle = 0$ [5], while the NADS are, in general, non-orthogonal due to Eq. (8): $\langle'\tilde{E}|\tilde{G}'\rangle \neq 0$. The NADS will be orthogonal if $COS(\theta/2)$ and $SIN(\theta/2)$ are real-valued functions, Eq. (8). It occurs when the nonadiabatic factors are removed. Thus, the nonadiabatic factors become related with the (non)orthogonality of the NADS. They cause rotation of the NADS in the Hilbert space, Fig.2. As the field nonadiabatic factors change in time, the direction, as well as the norm, of the NADS changes accordingly. Hence, the NADS form a kind of *dynamical non-orthogonal basis*. The ADS also result from a rotation in the Hilbert space (not shown in Fig.2) due to a unitary transformation on the BS [7] that, however, preserves the orthonormality of the states. A set of linearly independent state vectors can be made orthogonal (even orthonormal) following the Gram–Schmidt process. Here, the states will be preserved from procedures that are not physically justified because we consider that the "upgraded" states, *e.g.*, NADS, acquire given changes due to real physical processes acting on the quantum system. Any non-physical procedure, say, forcing the state to be orthogonal, may result in distortion, or even loss, of some physical properties and real physical effects in the quantum system. Instead, we will give a physical significance to the non-orhogonality of the NADS.

The non-orthogonality of the eigenstates is acceptable in modern quantum physics, where the concept of the observable is considered in a wider sense. Instead of "sharp" (von Neumann) observables, "smeared" observables associated with non-orthogonal eigenstates are considered [19]. The non-orthogonality of the NADS, however, requires a higher degree of caution in the interpretation of the results.

The general theory of quantum transitions, usually performed within the BS basis, assumes the following scenario. The quantum system, subject to a perturbation $\hat{H}'$, evolves from an initial BS $|i\rangle$ to, in general, a superposition state $|\psi\rangle$ that obeys the time dependent Schrödinger equation. The state $|\psi\rangle$ may not be an eigenstate of the Hamiltonian of the quantum system but it can always be expanded in terms of the BS, $|\psi\rangle = \sum_{j=1}^{n} c_j(t)|j\rangle$. Once the perturbation is terminated, the quantum system can be found in a final BS $|f\rangle$, if a measurement is performed on it. Thus, the quantum system undergoes a transition from the initial state $|i\rangle$ to some final (generally different) state $|f\rangle$. The transition probability is given by the square modulus of the projection of $|\psi\rangle$ onto $|f\rangle$: $P_{i \to f} = |\langle f|\psi\rangle|^2 = |c_f(t)|^2$. Although any solution of the Schrödinger equation can be expanded in terms of the BS, the projection on the BS cannot be fulfilled outside the zero-field region because the BS are not the states in which the system can be found in that region - the BS are not eigenstates of the adiabatic $\hat{H}_A$ or the non-adiabatic $\hat{H}_{NA}$ (*e.g.*, Eq.(1)) Hamiltonians that operate in the region of validity of ADS and NADS, respectively, Fig.2. Hence, the formal theory of quantum transitions is a kind of asymptotic theory. It gives the transition probability after the perturbation has terminated. The most important region for the quantum transitions, however, is that one where the interaction is substantially nonadiabatic, the shaded area in Fig.2. In that case, the conditions are even more severe: neither the BS nor the ADS but the NADS are the states in which the quantum system can be found in the region of nonadiabatic interactions.

In this work, we consider *an alternative approach* to the quantum transitions following strictly the physical idea in the adiabatic theorem: starting from given initial BS, the quantum system remains in that (but continuously modified) eigenstate of the instantaneous Hamiltonian if the perturbation/field evolves adiabatically. The quantum transitions *due to field* occur in the time region of nonadiabatic evolution. Thus, the following evolution sequences take place at ground state $|g\rangle \to |G\rangle \to |\tilde{G}'\rangle \Rightarrow |\tilde{E}'\rangle$ and excited state $|e\rangle \to |E\rangle \to |\tilde{E}'\rangle \Rightarrow |\tilde{G}'\rangle$ initial conditions, where the single line arrow "$\to$" shows adiabatic following process, and the hollow arrow "$\Rightarrow$" shows nonadiabatic transition. The quantum transitions *due to damping* occur continuously in time because of the constant damping rates.

In accordance with the above considerations, in order to obtain the transition probability in the region of nonadiabatic interactions, the projection of the initial NADS, *e.g.*, $|\tilde{G}'\rangle$, onto the final NADS, *e.g.*, $|\tilde{E}'\rangle$, must be found. Thus, the normalized matrix element (8) determines *the probability of the quantum transitions between the NADS*:

$$P_{|\tilde{G}'\rangle \to |\tilde{E}'\rangle} \equiv \frac{|\langle'\tilde{E}|\tilde{G}'\rangle|^2}{\langle'\tilde{G}|\tilde{G}'\rangle\langle'\tilde{E}|\tilde{E}'\rangle} = \frac{|SIN(\theta/2)(COS(\theta/2))^* - (SIN(\theta/2))^* COS(\theta/2)|^2}{(|SIN(\theta/2)|^2 + |COS(\theta/2)|^2)^2} \quad (9)$$

The exponential factors in Eqs.(6)-(8) exactly cancel each other out in (9). The expression of the transition probability (9) includes explicitly the contribution of the nonadiabatic factors through $SIN(\theta/2)$ and $COS(\theta/2)$ (that play role of the coefficients $c_j$ for the BS basis). It allows to explore the mechanism of the quantum transitions in the region of nonadiabatic interactions, where the transitions actually take place. As it can be seen from (9), the transition probability is nonzero in the presence of nonadiabatic factors, *i.e.*, at complex values of $SIN(\theta/2)$ and $COS(\theta/2)$, and becomes zero if the nonadiabatic factors are removed, *i.e.*, at real values of $SIN(\theta/2)$ and $COS(\theta/2)$. The latter proves the adiabatic theorem in terms of the NADS. Due to $\langle \tilde{G}' | \tilde{E}' \rangle = \langle \tilde{E}' | \tilde{G}' \rangle^*$, the condition of microreversibility holds, $P_{|\tilde{E}'\rangle \to |\tilde{G}'\rangle} = P_{|\tilde{G}'\rangle \to |\tilde{E}'\rangle}$.

One may distinguish two related reasons for quantum transitions: *a formal reason* and *a physical reason*. The formal reason for quantum transitions is the non-orthogonality of the NADS in the Hilbert space, Eq.(8). The physical reason for quantum transitions is the nonadiabatic factors acting on the quantum system, which, in fact, make the NADS non-orthogonal. Analyzing Eq.8, one may conclude that the nonadiabatic factors in the Hermitian part of the Hamiltonian (the field-matter interaction term $\hat{H}'$) leads to the same general effect as the nonadiabatic factors in the non-Hermitian part of the Hamiltonian (the damping term $\hat{H}_D$) – both factors make the NADS non-orthogonal thus leading to quantum transitions. Both reasons for quantum transitions are consistent with each other and, also, they are in agreement with the adiabatic theorem of quantum mechanics. All these make the NADS a natural physical basis for the consideration of quantum phenomena in the region of nonadiabatic interactions.

## 4. Conclusions

*In conclusion*, matrix elements from the nonadiabatic dressed states are found. The nonadiabatic dressed states form a complete set of linearly independent state vectors that are non-orthonormal in the Hilbert space. The latter is considered to be a formal reason for quantum transitions. The physical reason for the quantum transitions are the nonadiabatic factors due to external electromagnetic fields and damping caused by the environment. A closed form expression for the quantum transitions in the region of nonadiabatic interactions, where the transitions actually take place, is proposed for the first time based on the nonadiabatic dressed states.

## 5. Refernces